\documentclass{emulateapj}

\usepackage{txfonts}
\usepackage{graphicx}
\usepackage{natbib}
\usepackage{amssymb}
\bibpunct{(}{)}{;}{a}{}{,}
\newcommand{\sdss}{\emph{SDSS }}
\newcommand{\teff}{$T_{\rm eff}$}
\newcommand{\logg}{$\log{g}$}

\newcommand{\msun}{$M_\odot$}
\newcommand{\kms}{$\rm{km\,s^{-1}}$}
\newcommand{\masyr}{$\rm{mas\,yr^{-1}}$}
\newcommand{\M}{$[M/H]$}

\newcommand{\alfa}{$[\alpha/{\rm Fe}]$}

\shorttitle{Weighing the Galactic dark matter halo}
\shortauthors{Przybilla et al.}

\begin{document}


\title{\centering Weighing the Galactic dark matter halo: a lower mass
limit from the\\ fastest halo star known}


\author{Norbert Przybilla, Alfred Tillich and Ulrich Heber}
\affil{Dr. Karl Remeis-Observatory Bamberg \& ECAP, University Erlangen-Nuremberg, Sternwartstr. 7, D-96049 Bamberg, Germany}
\email{Norbert.Przybilla@sternwarte.uni-erlangen.de}

\and

\author{Ralf-Dieter Scholz}
\affil{Astrophysikalisches Institut Potsdam, An der Sternwarte 16,
D-14482 Potsdam, Germany}



\begin{abstract}
The mass of the Galactic dark matter halo is under vivid discussion. 
A recent study by Xue et al. (2008, ApJ, 684, 1143) revised the
Galactic halo mass downward by a factor of $\sim$2 relative to previous
work, based on the line-of-sight velocity distribution of $\sim$2400 
blue horizontal-branch (BHB) halo stars. The observations were
interpreted in a statistical approach using cosmological galaxy formation 
simulations, as only four of the 6D phase-space coordinates were determined. 
Here we concentrate on a close
investigation of the stars with highest negative radial velocity from
that sample. For one star, SDSSJ153935.67+023909.8 (J1539+0239 for
short), we succeed in measuring a significant proper motion, i.e.
full phase-space information is obtained. We confirm the star to be a
Population~II BHB star from an independent quantitative analysis 
of the {\em SDSS} spectrum -- providing the first NLTE study of any halo
BHB star -- and reconstruct its
3D trajectory in the Galactic potential. J1539+0239 turns out as the
fastest halo star known to date, with a Galactic rest-frame velocity
of 694$^{+300}_{-221}$\,\kms\ (full uncertainty range from Monte
Carlo error propagation) at its current position. The extreme kinematics of
the star allows a significant lower limit to be put on the halo mass
in order to keep it bound, of $M_{\rm halo}\ge1.7_{-1.1}^{+2.3}\times10^{12}$\,\msun.
We conclude that the Xue et al. results tend to underestimate the true
halo mass as their most likely mass
value is consistent with our analysis only at a level of 4\%.
However, our result confirms other studies that 
make use of the full phase-space information.\\
\end{abstract}


\keywords{dark matter --- Galaxy: halo --- stars: atmospheres --- stars:
horizontal-branch --- stars: kinematics and dynamics --- stars: Population II}



\section{Introduction}\label{sec:intro}
Knowledge of the properties of dark matter halos is an important issue
for our understanding of galaxy formation and evolution, and for
unveiling the nature of dark matter. 
The halo of the Milky Way therefore is of highest interest, as it allows unique 
observational constraints to be obtained for testing theoretical
models \citep[e.g.][]{1996ApJ...462..563N}. Several  
observational campaigns -- e.g. the Sloan Digital Sky Survey
\citep[\sdss,][]{2000AJ....120.1579Y}, the RAdial Velocity Experiment
\citep[\emph{RAVE},][]{2006AJ....132.1645S}, the Sloan Extension for Galactic
Understanding and Exploration \citep[\emph{SEGUE},][]{2009AJ....137.4377Y}  -- 
provide the tracers for studying halo properties, like the total mass of the halo 
and its extent.

Several studies in the past decade determined the halo mass from
ever increasing samples of halo stars, globular clusters and/or satellite 
galaxies. However, it is in fact only a few objects at highest velocity
that are affecting largely the mass estimates \citep{2003A&A...397..899S,2007MNRAS.379..755S}.
While larger halo masses of about $2\times10^{12}$\,\msun\ 
were favoured earlier 
\citep{1999MNRAS.310..645W,2003A&A...397..899S},
lower masses of about half this value were derived more recently 
\citep{2005MNRAS.364..433B,2007MNRAS.379..755S,2008ApJ...684.1143X}.  
The precise value determines, e.g. among others, whether satellite galaxies like 
the Magellanic Clouds \citep[e.g.][]{2006ApJ...652.1213K,2009AJ....137.4339C} or
hyper-velocity star\footnote{Hyper-velocity stars move at such high velocity
that they may be gravitationally unbound to the Galaxy. Their supposed
place of origin is the Galactic center, where they may have been
accelerated by gravitational interactions with the super-massive black
hole \citep{hills88}.} candidates \citep{2009ApJ...691L..63A} are on bound orbits, or not.

Most of the previous studies had to rely substantially on the
{\em distributions} of radial velocities in their samples to derive their conclusions as full 
space motions for halo objects are unavailable in many cases at present.
In such cases only four coordinates (i.e. two position values,
distance and radial velocity, RV) of the 6D phase space are
determined and the missing data (the proper motion components) are
handled in a statistical approach.
E.g., radial velocities for more than 10,000 blue halo stars from the \sdss were
measured in the yet most extensive study by \citet{2008ApJ...684.1143X}. 
The sample was composed of blue horizontal-branch, blue straggler, 
and main-sequence stars with effective temperatures roughly between 7,000 and 
10,000\,K according to their colours. Here, we focus on the fastest
stars of the Xue et al. sample in terms of {\em negative} line-of-sight velocity, 
indicating a {\em bound} orbit.

For one of them we were able to measure a significant proper motion,
which allowed a detailed 3D kinematic investigation when combined with a
quantitative spectroscopic analysis that facilitated the determination
of the star's distance. \object{SDSSJ153935.67+023909.8} (J1539+0239
for short) is an inbound\footnote{An extragalactic origin of the star as
an unbound low-mass hyper-velocity star from another galaxy is 
imaginable but unlikely. The local volume is devoid of galaxies in a 
wide range (several 10\,$\deg$)
around the infall direction of J1539+0239 at present
\citep[for the Local Group see][and later discoveries]{1999A&ARv...9..273V}. 
If ejected early 
after its formation, the star could have travelled $\sim$8\,Mpc during its
lifetime, such that a final answer on this may be only obtained when
full space motions of the nearby galaxies are known.} Population~II horizontal branch star with a
Galactic rest-frame (GRF) velocity of $\sim$700\,\kms\ at its current
position, making it the fastest halo object known. This allows 
a significant lower limit for the total halo mass of the Galaxy to be
set. The present work provides a glimpse to the detailed kinematic
investigations feasible once the European Space Agency's Gaia satellite mission 
\citep[e.g.][]{2005ESASP.576.....T}
becomes operational, at much higher precision.

\section{Target selection and proper motion}\label{sec:tar+pm}
\begin{figure}[t]
\begin{center}
\includegraphics[width=.915\linewidth]{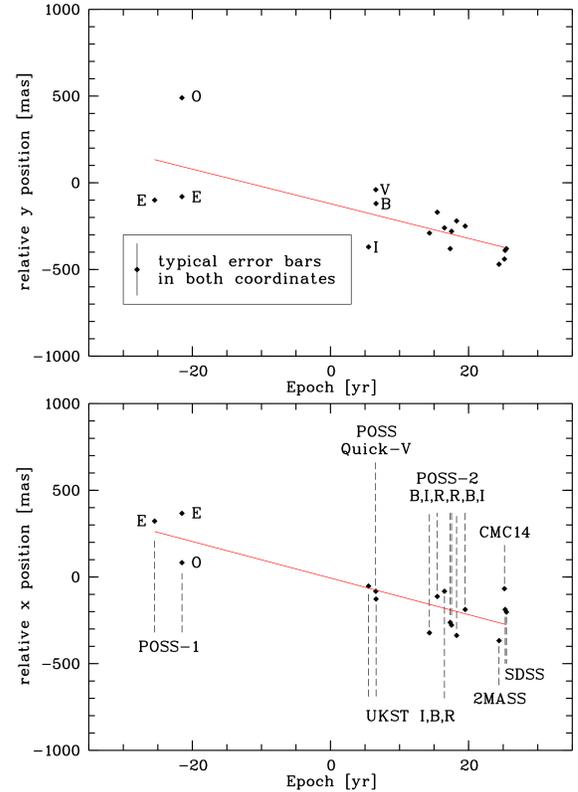}
\caption{\label{fig_PMfit}Linear fit of the position measurements for J1539+0239 with a zero epoch of 1975.39.}
\end{center}
\end{figure}

In a previous paper \citep{2009A&A...507L..37T} we already
investigated the high-velocity tail of the \cite{2008ApJ...684.1143X} sample 
and found a hyper-velocity candidate of spectral type A among the stars with highest 
{\em positive} radial velocities in the GRF. Here we study the most 
extreme stars approaching us, applying the same techniques. 
We selected all stars with GRF velocities $v_{\rm GRF}<-350$~\kms\ from the 
RV-based sample of \cite{2008ApJ...684.1143X} and obtained 5 
targets for which we attempted to measure proper motions. 
All available independent position measurements on Schmidt plates
(APM - \cite{2000yCat.1267....0M};
SSS - \cite{2001MNRAS.326.1279H}) were collected
and combined with the \sdss and other available positions
(CMC14~\cite{2006yCat.1304....0C};
2MASS - \cite{2003tmc..book.....C};
UKIDSS - \cite{2007MNRAS.379.1599L})
for a first linear proper motion fit. However, there were even
more measurements of Schmidt plates, from up to 14 different epochs in case
of overlapping plates of the Digitised Sky
Surveys\footnote{http://archive.stsci.edu/cgi-bin/dss\_plate\_finder}.
FITS images of 15 by 15 arcmin size were extracted from all available plates
and ESO MIDAS tools were used to measure positions. For this purpose,
we selected compact background galaxies around each target, identified from
\emph{SDSS}, to transform the target positions on all the Schmidt plates to the
\sdss system. The small fields allowed us to apply a simple model
(shift+rotation) and to achieve a higher accuracy in our final proper motion fit 
for all our targets (see e.g. Fig.~\ref{fig_PMfit}). 
We consider a positive detection if the proper motion errors 
$\sigma$ of both components are below 3.0\,\masyr\ and if at least
one of the components is significant (above 3$\sigma$). 
For J1539+0239, the brightest of our five targets, we found a highly significant 
proper motion of 
$\mu_\alpha\cos\,\delta=-$10.6$\pm$1.6\,\masyr\ and 
$\mu_\delta=-$10.0$\pm$2.3\,{\masyr}, 
whereas for the other four fainter stars (with $V$ magnitudes
between 17 and 20) the proper motion was consistent with zero.
The proper motions as well as their transformations from
equatorial to Galactic coordinates are summarized in Table~\ref{tab_HVS}.
 
\section{Observations and quantitative spectroscopy}\label{sec:spec}

In order to exclude RV variability, we re-observed J1539+0239 with 
the TWIN spectrograph at the 3.5m telescope on Calar Alto in  
May 2009. Radial velocities were derived by 
$\chi^2$-fitting of adequate synthetic spectra over the full spectral
range, yielding a heliocentric radial velocity of $v_{\rm
rad}=-366.6\pm4.0$\,\kms\ for the TWIN spectrum, which is consistent with the $v_{\rm
rad}=-372.6\pm5.8$\,\kms\ measured from the \sdss data within the mutual uncertainties. 
We use the latter value for the kinematic study. 

A quantitative analysis of J1539+0239 was carried
out following the hybrid NLTE approach discussed by
\cite{2006A&A...445.1099P}. In brief,
line-blanketed LTE model atmospheres were computed with
ATLAS9 \citep{1993KurCD..13.....K} and NLTE (and LTE) line-formation
calculations were performed using updated versions of {\sc Detail}
and {\sc Surface} \citep{1981PhDT.......113G,bugi85}. Many
astrophysically important chemical species were treated in
NLTE, using state-of-the-art model atoms (H: Przybilla \& Butler~2004a; 
C\,{\sc i}: Przybilla et al.~2001b; N\,{\sc i}: Przybilla \& Butler~2001;
O\,{\sc i}: Przybilla et al.~2000; Mg\,{\sc i/ii}: Przybilla et
al.~2001a; Ti\,{\sc ii} and Fe\,{\sc ii}: Becker~1998).

\begin{figure*}
\begin{center}
\resizebox{.99\hsize}{!}{\includegraphics{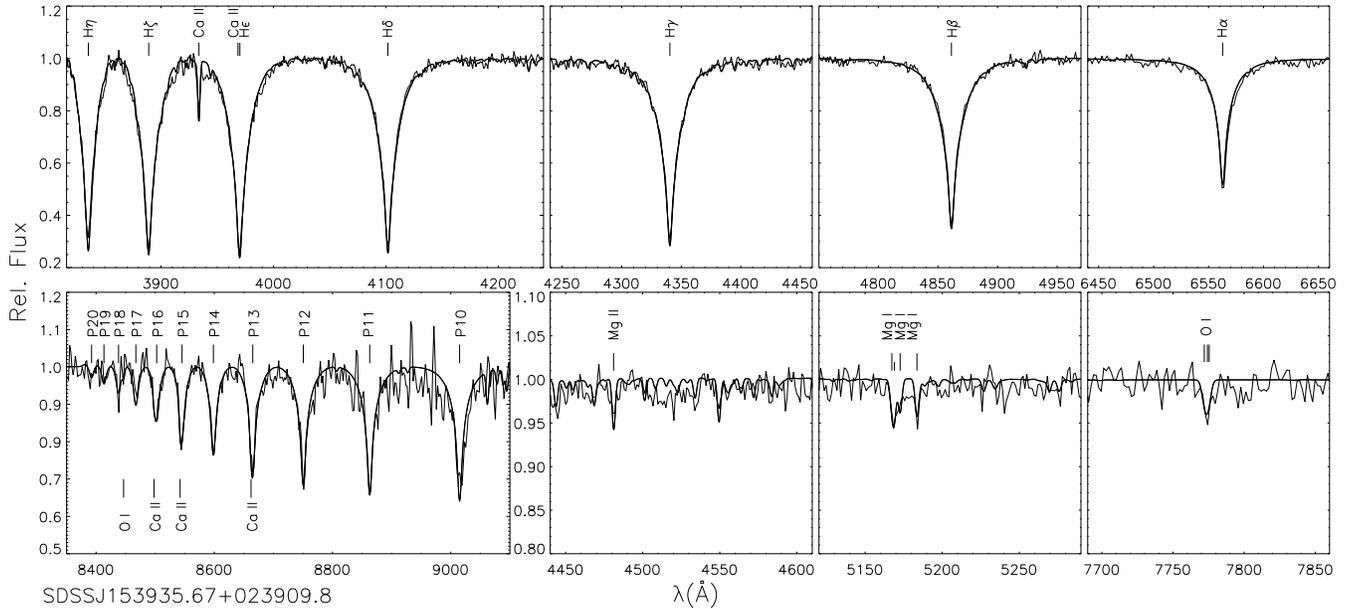}}\\[1.95mm]
\caption{\label{linefits}
Comparison of NLTE spectrum synthesis (thick line) with observation 
(\emph{SDSS}, thin wiggly line) for J1539+0239. 
Displayed are selected
regions around the Balmer lines, the higher Paschen series,
Mg {\sc ii}\,$\lambda$4481{\AA}, the Mg {\sc i}\,b and the near-IR
O {\sc i} triplets.}
\end{center}
\end{figure*}

The effective temperature {\teff} and the surface gravity
$\log g$ were determined by fits to the Stark-broadened Balmer and Paschen
lines and an ionization equilibrium, here of Mg {\sc i/ii}, in analogy to 
previous work on hyper-velocity stars at similar temperatures
\citep{2008A&A...488L..51P,2009A&A...507L..37T}. 
The stellar metallicity was derived by model fits to the observed metal line
spectra. Results are listed in Table~\ref{tab_HVS} and a
comparison of the resulting final synthetic spectrum with
observation is shown in Fig.~\ref{linefits}.
Overall, excellent agreement is obtained for the strategic spectral lines
throughout the entire wavelength range. Our stellar parameters 
(\teff$ =7700\pm250$\,K, $\log g=3.00\pm0.15$) 
are consistent with those derived in the LTE analysis by
\citet{2008ApJ...684.1143X}: \teff$ =7807$\,K, $\log g=3.16$.
We constrained the errors in the stellar parameters by the quality of
the match of the spectral indicators within the given S/N limitations.

Its parameters place J1539+0239 on the horizontal branch 
at a mass of 0.68$\pm$0.05\,\msun\ as
derived by comparing the position of the star in the \teff--$\log
g$-diagram to predictions of the evolutionary models of \cite{1993ApJ...419..596D}. 
No rotational broadening was detected at the resolution of the \sdss spectrum. 
The metallicity is lower than solar by
a factor of 100 and the abundances of the $\alpha$-elements 
are enhanced by about 0.4\,dex with respect to iron, which is typical for the halo 
population. 
We conclude that the star is a horizontal branch star of Population~II. 
All results are summarised in Table \ref{tab_HVS}.

\begin{table}
\caption{Results of the spectroscopic and kinematic analysis of J1539+0239.}
\label{tab_HVS}
\begin{center}
\begin{tabular}{lc|lc}
\hline\hline
$V$\,(mag)$^{\rm a} $     & 15.72 $\pm$ 0.02 & 
$E(B-V)$\,(mag)$^{\rm b} $ &  0.04 $\pm$ 0.03 \\
$\mu_\alpha\cos\delta$\,(\masyr) &   $-$10.6 $\pm$ 1.6 & 
$\mu_\delta$\,(\masyr) &   $-$10.0 $\pm$ 2.3 \\
$l$\,(deg) & 8.9836 & $b$\,(deg) & +42.9515\\
$\mu_l\cos b$\,(\masyr) & $-$14.3 $\pm$ 2.1 &
$\mu_b$\,(\masyr) & $+$2.8 $\pm$ 1.9\\
\teff\,(K)      &  7700 $\pm$ 250  & 
 \logg (cgs)      &  3.00 $\pm$ 0.15  \\
 \M          &  $-$2.0$\alpha$        &  
 \alfa        &  +0.4 \\  
 $M/M_\odot$ &  0.68 $\pm$ 0.05  &  
 $d$\,(kpc)    & 12.0 $\pm$ 2.3    \\
 $v_{\rm rad}$ (\kms)   & $-$372.6 $\pm$ 5.8 &
 $v_{\rm GRF}$\,(\kms) & 694 $^{+300}_{-221}$ \\
 v$_{\rm esc}$\,(\kms) & 519 && \\
 \hline  
 \multicolumn{4}{l}{\small $^{\rm a}$ The visual magnitude has been
derived from \sdss magnitudes}\\
\multicolumn{4}{l}{\small following \cite{2006A&A...460..339J}}\\
 \multicolumn{4}{l}{\small $^{\rm b}$ The interstellar colour excess
E(B-V) has been determined by}\\
\multicolumn{4}{l}{\small comparing the
observed colours to synthetic ones from the model}\\
\multicolumn{4}{l}{\small spectral energy distribution. }\\

\end{tabular}
\end{center}
\end{table}

\begin{figure}
\begin{center}
\resizebox{.85\hsize}{!}{\includegraphics{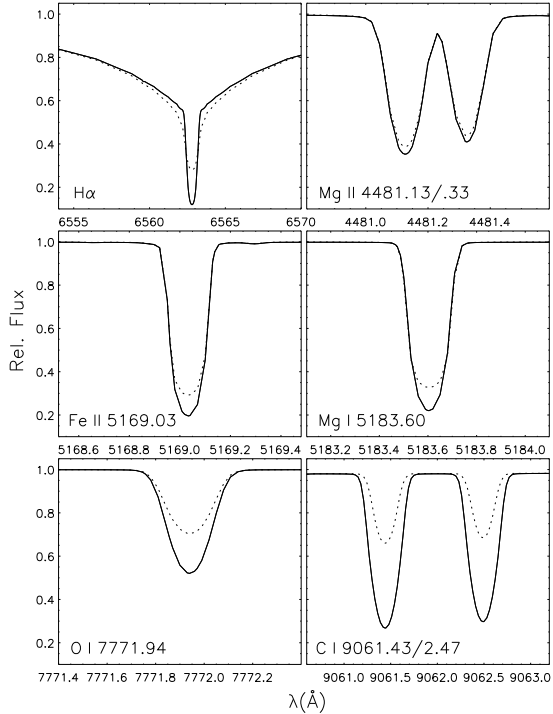}}\\[1.95mm]
\caption{\label{theo_lines}
Comparison of NLTE (full lines) and LTE (dotted) line profiles
from calculations for our adopted model parameters for J1539+0239. 
Displayed are examples of strong lines that can be used for diagnostics at the
\emph{SDSS} spectral resolution.}
\end{center}
\end{figure}

Before proceeding to the further discussion it
may be instructive to take a closer look on the
spectrum synthesis in NLTE and LTE, as this has not been done
for Population\,II blue horizontal branch-stars so far. A few comparisons of NLTE
and LTE profiles are therefore shown in Fig.~\ref{theo_lines}.
The combination of a higher {\teff} than typically found for
Population\,II stars and the diminished line blocking because of the
low metal content results in a hardened radiation field which, along
with reduced thermalizing effects because of smaller collision rates in the 
low-density atmosphere, leads to
pronounced NLTE effects on many diagnostic lines. NLTE strengthening
is found for the Doppler core of H$\alpha$ -- in line with the
behaviour in cool stars \citep[e.g.][]{2004ApJ...610L..61P} --, while the
inner line wings are weakened. The higher Balmer and Paschen lines show much lower
deviations from LTE. Our calculations predict the majority of the weak
lines to be described well by the assumption of LTE. On the other hand,
many of the stronger metal lines, as e.g. of C\,{\sc i},
O\,{\sc i}, Mg\,{\sc i/ii} or Fe\,{\sc ii} -- the diagnostic lines at
the spectroscopic resolution achieved within the {\em SDSS} -- show pronounced NLTE
strengthening. In order to reproduce
the NLTE equivalent widths of these particular lines, abundance
corrections need to be applied of about 0.1\,dex (Mg\,{\sc ii}),
0.2\,dex (Mg\,{\sc i}, Fe\,{\sc ii}), 0.5\,dex (O\,{\sc i}) and 
1.9\,dex (C\,{\sc i}) in LTE. Note, however,
that these line are close to saturation. LTE computations with
increased abundances can therefore not reproduce the NLTE line depths at all,
instead stronger line wings develop. Abundance studies based on equivalent
widths may therefore be misleading, as such differences remain
unnoticed. An investigation at high spectral resolution would therefore
be worthwhile in order to facilitate the NLTE effects to be studied in
detail.

\section{Distance, kinematics and errors}\label{sec:kine}
Using the mass, effective temperature, gravity and extinction-corrected apparent
magnitude we derive the distance following \cite{2001A&A...378..907R} 
using the fluxes from the final model spectrum in analogy to previous 
work \citep{2009A&A...507L..37T}. The distance error is dominated by the 
gravity error. 

\begin{figure}
\begin{center}
\includegraphics[width=.995\linewidth]{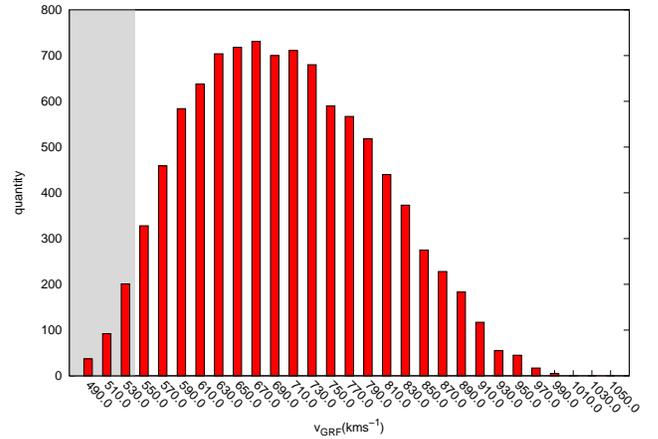}
\caption{\label{fig:veldistrib_J1539}   
Galactic restframe velocity distribution for J1539+0239 derived from a Monte 
Carlo procedure with a depth of 10,000. The grey shaded area
indicates the fraction of trajectories that would be bound for 
$M_{\rm halo}=1.0\times10^{12}$\,\msun, the most likely halo mass of
\citet{2008ApJ...684.1143X}.
}
\end{center}
\end{figure}

Applying the Galactic potential of
\citet{1991RMxAA..22..255A} we calculated orbits and 
reconstructed the path of the star through the Galactic halo with the program of
\citet{1992AN....313...69O}. 
The distance of the Galactic center from the Sun was adopted to be 8.0\,kpc and
the Sun's motion with respect to the local standard of rest was taken from 
\citet{1998MNRAS.298..387D}. 
As the RV is well known the error of the space motion is dominated
by that of the distance, ruled by the gravity error, and those of the 
proper
motion components. Varying these three quantities within their respective errors 
we applied a Monte-Carlo procedure to derive the median GRF velocity
$v_{\rm GRF}$ at the present 
location and the velocity distribution (see Fig.~\ref{fig:veldistrib_J1539}) and
compare with the local escape velocity $v_{\rm esc}$ as calculated from the 
Galactic potential of \citet{1991RMxAA..22..255A}.  
 
A median GRF velocity of 694\,\kms\ (with the velocity distribution ranging 
from $-221$ to +300\,\kms\ around this value) makes J1539+0239 the
fastest halo star known, superseding \object{CS~22183$-$0014}
\citep[at $v_{\rm GRF}$\,=\,635$\pm$127\,\kms,][]{2003A&A...397..899S}. This value 
is above the local escape velocity of $v_{\rm esc}\approx519$\,\kms\ in the 
potential of \citet{1991RMxAA..22..255A}. This three component potential 
consists of a central bulge and a disk, which have a combined mass of 
$M_{\rm bulge+disk}\approx10^{11}$\,\msun. The halo out to 100\,kpc is assumed to have a total mass of 
$M_{\rm halo}\approx8\times10^{11}$\,\msun. 
This total halo mass is insufficient to keep the star bound to the Galaxy. 

\begin{figure}[t]
\begin{center}
\includegraphics[width=\linewidth]{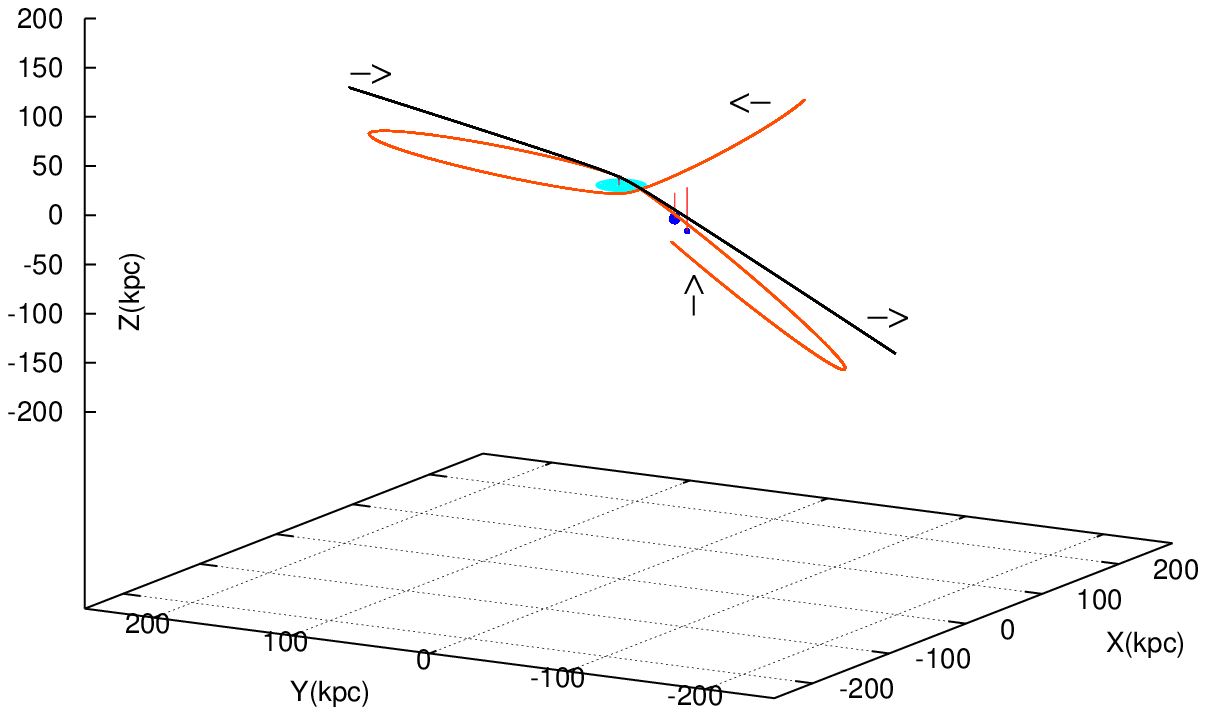}\\
\includegraphics[width=\linewidth]{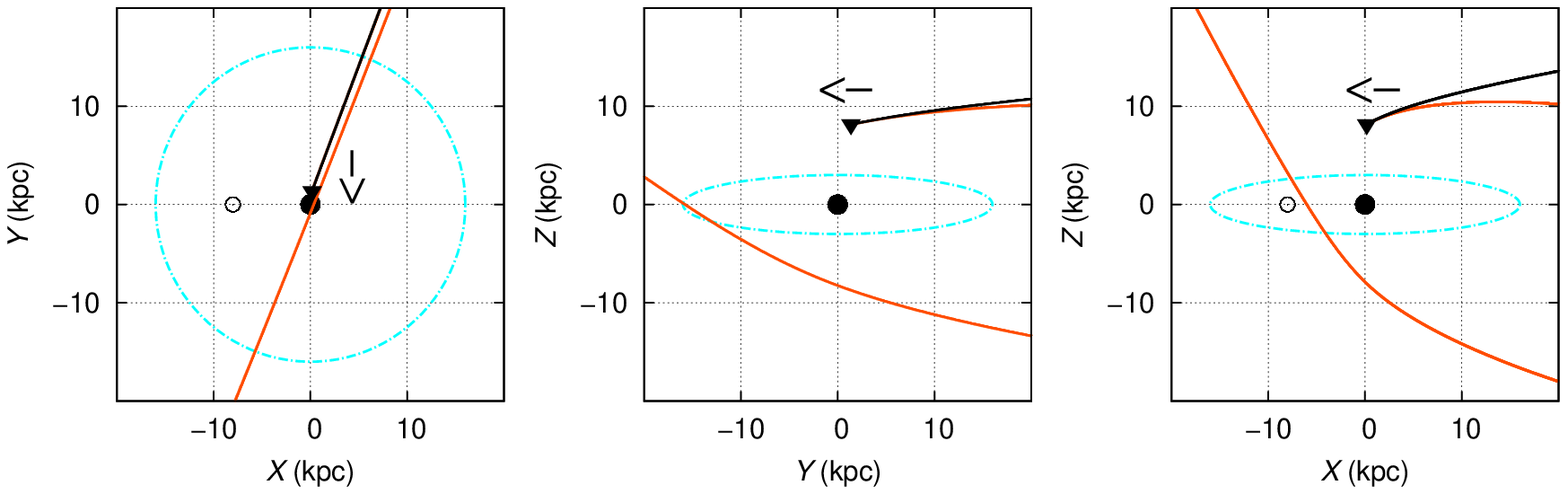}
\caption{\label{fig_J1539_3dplot}
Upper panel: Trajectories for the metal-poor horizontal branch star J1539+0239, relatively
to the Galactic disk (light blue). Applying a standard potential 
\citep{1991RMxAA..22..255A} the trajectory is unbound (black,
$t\approx\pm0.5$\,Gyr), while increasing the halo density we found a 
bound trajectory (red, $-$3\,Gyr\,$\lesssim$\,$t$\,$\lesssim$\,+2\,Gyr). 
For reference, the Magellanic Clouds (blue dots) and the current position of
J1539+0239 (black triangle) are marked. Lower panel: schematic 2D-visualization of the 
the central region. Details of the trajectory in the immediate past
and around the previous pericenter passage of J1539+0239 (bound solution) are shown
in the $X$--$Y$, $Y$--$Z$ and $X$--$Z$ planes. The positions of the
Galactic center (black dot) and of the Sun are indicated.}
\end{center}
\end{figure}

J1539+0239 is located in the northern Galactic hemisphere ($l\approx9.0$,
$b\approx42.95$), in the direction close to the Sagittarius stream
\citep{2006ApJ...642L.137B,2006ApJ...651..167F} and in particular
close to the 
globular clusters \object{NGC 5904} and \object{Palomar~5}, with its
tidal tail \citep{2003AJ...126.2385O}. However, it is unlikely that
J1539+0239 is related to either of those because of its distinct
space motion and its position in the foreground. 
Its position and kinematics also rule out a scenario of a {\em
recent} acceleration by an interaction with the super-massive black hole in the 
Galactic center, the principal mechanism for generating hyper-velocity 
stars \citep{hills88}.
The star is currently {\em approaching} the Galactic disk and will 
pass the Galactic center at a minimum distance of about 8\,kpc in the
future. See Fig.~\ref{fig_J1539_3dplot} for a
visualization of the orbit of J1539+0239 in the Galactic halo (upper
panel) and a for a magnification of the Galactic
disk region of the trajectory (lower panel).

Hence, in order to keep the trajectory of J1539+0239 bound to the Galaxy, 
the dark matter halo mass needs to be adjusted. We carried out numerical 
experiments increasing the halo density by a constant factor. Finally we 
found a bound trajectory for a minimum mass of 
$M_{\rm halo}^{\rm new}=1.7\times10^{12}$\,\msun. 
The last pericenter passage occured at a distance of $\sim$7.7\,kpc and
the apocenter distance of the star's trajectory is located far out 
in the halo, at $\sim$250\,kpc in this case.
If we take into account the full velocity distribution (see 
Fig.~\ref{fig:veldistrib_J1539}) of the star we even can derive solutions 
for the extrema, which correspond to the absolute errors, giving
$M_{\rm halo}\sim1.7_{-1.1}^{+2.3}\times10^{12}$\,\msun.

Whether the star is bound to the Galaxy highly depends on the Galactic
potential adopted, in particular on the mass of the dark matter halo, as pointed 
out by \citet{2009ApJ...691L..63A}. Our $M_{\rm halo}$ is similar to
values found in several recent studies. 
\citet{1999MNRAS.310..645W} used 27 satellite galaxies and globular clusters, 
by assuming that they are bound and derived a total Galactic halo mass of 
$M_{\rm halo}\sim1.9_{-1.7}^{+3.6}\times10^{12}$\,\msun. 
This value matches our derivation but has a larger uncertainty. 
\citet{2003A&A...397..899S} used 11 satellite galaxies, 137 globular clusters 
and 413 field horizontal branch stars to derive a total Galactic mass. 
The exclusion of \object{Leo~I} from
their sample would lower the total Galactic 
mass from $M_{\rm total}\sim2.5_{-1.0}^{+0.5}\times10^{12}$\,\msun\ to a value of 
$M_{\rm total}\sim1.8_{-0.7}^{+0.4}\times10^{12}$\,\msun, both again in
excellent agreement with our finding. 

On the other hand, less than 4\% of
the trajectories resulting from our MC simulations would be bound for
the most likely mass of Xue et al., $M_{\rm
halo}=1.0\times10^{12}$\,\msun\ (grey shaded area in
Fig.~\ref{fig:veldistrib_J1539}). Their RV study of $\sim$2400 blue
horizontal-branch stars -- which includes J1539+0239 but lacks any
proper motion measurements -- therefore likely
underestimate the Galactic halo mass.


\section{Summary and Conclusions}\label{sec:conclusion} 

We reported the quantitative spectral analysis of a high-velocity star
from the sample of faint blue halo stars of \citet{2008ApJ...684.1143X}.
J1539+0239 was confirmed to be a Population~II blue horizontal branch star
with a low metallicity of $[$Fe/H$]=-2.0$ and the characteristic enhancement 
of $\alpha$-elements. Hereby, we performed a NLTE analysis of a
halo BHB star for the first time. While the majority of the weak lines were
confirmed to be formed close to LTE conditions, many of the stronger
metal lines -- which are important diagnostics at the spectral
resolution achieved within the {\em SDSS} -- show pronounced NLTE
strengthening, with the differences between the derived LTE and NLTE
abundances amounting to 0.1\,dex to 0.5\,dex typically. In addition
to information on the chemical composition, the
radial velocity, proper motion and spectroscopic distance were derived and
a detailed kinematical analysis was performed. 

Carrying out kinematical numerical experiments using the Galactic
potential of \citet{1991RMxAA..22..255A} in order to obtain an 
orbit of J1539+0239 gravitationally bound to the Milky Way, we found that the mass of the 
dark halo has to be at least $M_{\rm halo}\sim1.7_{-1.1}^{+2.3}\times10^{12}$\,\msun
(absolute uncertainties from extrema in MC error propagation) .
This mass limit is in good agreement with several previous studies
\citep{1999MNRAS.310..645W,2003A&A...397..899S,2009ApJ...691L..63A}.
However, the significantly lower most likely mass value of
\cite{2008ApJ...684.1143X} is consistent with our analysis only
at a level of 4\%, i.e. it likely underestimates the Galactic dark halo
mass.

We conclude, that if the kinematics of a halo star is extraordinary enough, 
and the errors within the analysis are small, even {\em one} star alone can
provide a significant lower limit to the dark matter halo mass, and to
the total mass of the Milky Way (halo\,+\,bulge\,+\,disk), 
here $M_{\rm total}\ge1.8_{-1.1}^{+2.3}\times10^{12}$\,\msun. 
The determining factor is that full kinematic information is
available, as it will become routine in the era of the Gaia space
mission, at much higher precision.

\acknowledgments
A.T. acknowledges funding by the Deutsche Forschungsgemeinschaft (DFG) through  
grant HE1356/45-1. Travel to the DSAZ (Calar Alto, Spain) was
supported by the DFG under grant HE1356/50-1.
We are very grateful to Stephan Geier for stimulating discussions and advice.
Our thanks go to S. M\"{u}ller and T. Kupfer for observing and reducing the 
data from DSAZ.
Funding for the \sdss and {\em SDSS}-II has been provided by the Alfred P.
Sloan Foundation, the Participating Institutions, the National Science
Foundation, the U.S. Department of Energy, the National Aeronautics
and Space Administration, the Japanese Monbukagakusho, the Max Planck
Society, and the Higher Education Funding Council for England. The
\sdss Web Site is http://www.sdss.org/.\\



{\it Facilities:} \facility{Sloan}, \facility{CAO:3.5m (TWIN)}.

\end{document}